\documentclass{appolb}
\usepackage{graphicx}
\usepackage{amsmath}
\usepackage{amssymb}

\begin{document}
\title{\LARGE\bf Bhabha scattering at future colliders\vspace{1mm}\\ with BHLUMI/BHWIDE%
\thanks{Presented by WP at the XLVI International Conference of Theoretical Physics ``Matter To The Deepest'' (MTTD),
 Katowice, Poland, 15--19 September 2025.}%
\vspace{8mm}
}
\author{Wies{\l}aw P{\l}aczek
\address{Faculty of Physics, Astronomy and Applied Computer Science, \\
Jagiellonian University, ul.\ {\L}ojasiewicza 11, 30-348 Krakow, Poland}
\\[3mm]
{Maciej Skrzypek 
\address{Institute of Nuclear Physics, Polish Academy of Sciences, \\ 31-342 Krak\'ow,  
ul. Radzikowskiego 152, Poland}
}
\\[3mm]
{Bennie F.L.\ Ward 
\address{Department of Physics, Baylor University, \\
One Bear Place \# 97316, Waco, TX 76798-7316, USA}
}
\\[3mm]
{Scott A.\ Yost
\address{Department of Physics, The Citadel, \\
171 Moultrie Street, Charleston, SC 29409, USA}
}
\\[3mm]
}
\maketitle
\begin{abstract}
\begin{center}
{\bf Abstract}
\end{center}
\noindent
In this paper,  we briefly present the Monte Carlo event generators {\sf BHLUMI} and {\sf BHWIDE} for small
and large angle Bhabha scattering, respectively, and discuss possible ways of their improvements
in order to satisfy precision needs of future electron--positron colliders.
\end{abstract}

\newpage
%================  
\section{Introduction}
\label{sec:Intro}
%================

Several future collider projects are being considered by the high-energy physics (HEP) community for the post-LHC era
of particle physics research. Since no signals of physics beyond the Standard Model (BSM) have been observed so far at 
the LHC, in the 2020 update of European Strategy for Particle Physics (ESPP) an electron--positron collider was indicated
as the preferred choice to study the Higgs boson in greater detail as well as to perform precision measurements that would possibly
reveal deviations from the Standard Model (SM) predictions, to be further studied in a next generation 
higher-energy (electron--positron, hadron or muon) collider \cite{EuropeanStrategyforParticlePhysicsPreparatoryGroup:2019qin}.
The possible options for the $e^+e^-$ collider discussed then included an International Linear Collider (ILC) in Japan \cite{ILC},
a Compact Linear Collider (CLIC) \cite{CLIC} or Future Circular Collider (FCC-ee) \cite{FCC-ee} at CERN and
a Circular Electron Positron Collider (CEPC) in China \cite{CEPC}.
The aim of the ongoing process of the 2026 update of ESPP is to choose the preferred option for CERN. 
In addition to the aforementioned colliders, three other projects have been taken into considerations: LEP3 \cite{Anastopoulos:2025jyh}, 
Linear Collider Factory (LCF) \cite{LinearCollider:2025lya} and LHeC \cite{LHeCStudyGroup:2012zhm}.
At the moment, FCC-ee seems to be the preferred option as the next HEP collider for CERN \cite{deBlas:2025gyz}.

FCC-ee is planned to run in a few stages: at the $Z$-boson peak, at the $W^+W^-$ production threshold, at the maximum
of the $ZH$ production cross-section, at the $t\bar{t}$ production threshold, and maybe at some more collision energy points.
Numbers of events to be collected at the $Z$ peak and at the $WW$ threshold are expected to exceed that of LEP experiments
by several orders of magnitude. This would allow to improve the LEP measurements of key electroweak observables 
by factors from 10 to 500 \cite{Proceedings:2019vxr,FCC:2018byv,FCC:2025lpp}.
In order to correctly interpret these measurements within SM or BSM, the experimental precision needs to be matched by
theoretical predictions at the same level or better.  

Among the processes of importance for experimental studies at electron--positron colliders is Bhabha scattering, i.e.\ 
the process $e^+e^- \rightarrow e^+e^-$. It is usually divided into two classes depending on the  range 
of an electron/positron scattering angle:
(1) small-angle Bhabha scattering (SABS), typically with $\theta_e \lesssim 100\,$mrad, and 
(2) large-angle Bhabha scattering (LABS),  typically with $\theta_e \gtrsim 100\,$mrad. 
Theoretical predictions for these processes that could be useful for experimental analyses should be provided in the form
of Monte Carlo event generators (MCEGs). 

In the following, we discuss two MCEGs -- developed originally for LEP -- in the context of future $e^+e^-$ collider needs
for theory predictions of the Bhabha scattering processes. Section~2 is devoted to 
{\sf BHLUMI} \cite{Jadach:1991by,Jadach:1996is,Bhlumi:code} for SABS
and Section~3 -- to {\sf BHWIDE} \cite{Jadach:1995nk,Bhwide:code} for LABS. In Section~4, we conclude our paper.    

%================  
\section{SABS with {\sf BHLUMI}}
\label{sec:Bhlumi}
%================
Integrated luminosity is a very important parameter of particle colliders. It is used to translate numbers of experimentally observed events
into cross sections of physical processes, 
\begin{equation}
\sigma = \frac{N}{\cal L}\,,
\label{eq:sigNL}
\end{equation}
where $\sigma$ denotes a cross section of some physical process, $N$ is a number of events observed in a particle detector
and ${\cal L}$ is the luminosity.
Then, the resulting cross section can be compared with predictions of theoretical models.   
Thus, precise knowledge of the collider luminosity is important for confronting experimental measurements with theory
-- in precision tests of the Standard Model (SM), on the one hand, and in searches for beyond the Standard Model (BSM) phenomena,
on the other hand.

One of the methods of the luminosity measurement is to identify some reference process for which, on the experimental side, 
high event statistics $N_{\rm ref}$ can be collected, with low background and good control of systematics, 
and, on the theory side, high-precision calculations of the corresponding cross section $\sigma_{\rm ref}$ 
are possible, with negligible new physics contributions, i.e.
\begin{equation}
{\cal L} = \frac{N_{\rm ref}}{\sigma_{\rm ref}}\,, \qquad\qquad 
\frac{\delta{\cal L}}{{\cal L}} = \frac{\delta N_{\rm ref}}{N_{\rm ref}} \oplus \frac{\delta\sigma_{\rm ref}}{\sigma_{\rm ref}}\,.
\label{eq:LerL}
\end{equation}
As can be seen from the second formula in Eq.~(\ref{eq:LerL}), the luminosity error consists of two contributions:
(1) an experimental error related to the measurement of $N_{\rm ref}$ and (2), a theoretical error resulting from 
accuracy of calculating $\sigma_{\rm ref}$. Generally, it is required that
$\delta\sigma_{\rm ref}/\sigma_{\rm ref} \lesssim \delta N_{\rm ref}/N_{\rm ref}$,
i.e.\ the luminosity measurement should not be limited by theory predictions.

At LEP, small-angle Bhabha scattering (SABS) was chosen as the main reference process for the luminosity measurement
\cite{Jadach:1996gu}.  It is dominated by the $t$-channel $\gamma$ exchange which, in principle, is a pure QED process,
i.e.\ high-precision theoretical calculations of $\sigma_{\rm ref}$ are possible. 
Because $d\sigma/d\theta_e \propto 1/\theta_e^3$, high event statistics can be collected by experiments at low angles,
$\theta_e\lesssim 100\,$mrad. Thanks to both experimental and theoretical efforts, the relative precision of the luminosity measurement
at LEP reached the level of $\sim 5\times 10^{-4}$ at the $Z$-boson peak \cite{ALEPH:2005ab}.
For the calculations of $\sigma_{\rm ref}$, all four LEP experiments used the MCEG {\sf BHLUMI~4.04} 
\cite{Jadach:1996is,Bhlumi:code}.

At future $e^+e^-$ colliders, the respective experimental precision can reach $\lesssim 10^{-4}$ at the $Z$ pole and ${\cal O}(10^{-3})$ 
at higher energies. This poses a big challenge for theoretical calculations. 
The main effects contributing to the error budget of luminometry at these colliders on the theory side and the ways to achieve 
the required precision were discussed in Refs.~\cite{Jadach:2018jjo,Jadach:2021ayv}.
There, the theoretical error for LEP at the $Z$ pole was reassessed down to $0.037\%$, 
and forecasts were made for achieving the precision 
of $10^{-4}$ at the $Z$ pole at FCC-ee and the sub-permil precision level at higher energies at this as well as other planned 
$e^+e^-$ colliders.   
These estimates were more recently updated in Refs.~\cite{Skrzypek:2024gku,Ward:2024frh}. 
In particular, the expected error for FCC-ee at the $Z$ pole was reduced to $0.7\times 10^{-4}$, and precision predictions for some
higher collision energies were also improved.

It was argued \cite{Jadach:1996gu} that when aiming at high numerical precision of the SABS cross section,
it is better to reorder a perturbative series in calculations of radiative corrections from the usual expansion in the QED coupling
constant $\alpha$ to the expansion in this coupling multiplied by the so-called big log, i.e.\ $\alpha^nL^m (m,n = 1,2, \ldots, m\leq n)$,
where $L\equiv \ln(|t|/m_e^2) - 1$. The latter was called the ``pragmatic'' QED expansion. In particular, it was shown that at LEP energies, 
the ${\cal O}(\alpha^2L^2)$ coefficient, being formally of the second order in the traditional expansion, is by a factor up to $6$
larger than the sub-leading first-order ${\cal O}(\alpha)$ factor.
Following this pragmatic expansion, it turned out that for LEP it was sufficient to include terms with the coefficients 
$\alpha L$, $\alpha$ and $\alpha^2L^2$. This was done in {\sf BHLUMI~4.04} and was called ${\cal O}(\alpha^2)_{\rm prag}$.
In order to reach the precision required for the future $e^+e^-$ colliders, in particular FCC-ee, this has to be upgraded to
${\cal O}(\alpha^3)_{\rm prag}$, i.e.\ additional terms proportional to the coefficients $\alpha^2L$ and $\alpha^3L^3$ need to be included.

The important feature of {\sf BHLUMI} is that it is based on the Yennie--Frautschi--Suura (YFS) exclusive exponentiation (EEX) method
\cite{Yennie:1961ad} in which all the infrared (IR) singularities are summed up to the infinite order and canceled properly 
in the so-called YFS form factor.  Then, calculated perturbatively IR-finite residuals  feature faster convergence than 
radiative corrections evaluated in the standard order-by-order approach. An important merit of {\sf BHLUMI}  is also 
a dedicated efficient Monte Carlo (MC) algorithm which features, among others, the exact multiphoton phase space.      

We have a three-stage plan of improvements in {\sf BHLUMI} that would allow to reach the luminosity precision goals set up 
by the future $e^+e^-$ colliders.
\vspace{-3mm}
\begin{description}
\item{\bf Stage~1:} 
The mentioned above photonic ${\cal O}(\alpha^2L)$ and ${\cal O}(\alpha^3L^3)$ corrections needed for ${\cal O}(\alpha^3)_{\rm prag}$ 
predictions are already available and can be implemented in {\sf BHLUMI}. The ${\cal O}(\alpha^3L^3)$ corrections are included 
in the leading-log Monte Carlo program {\sf LUMLOG} which is part of the {\sf BHLUMI~4.04} package. 
The ${\cal O}(\alpha^2L)$ corrections were calculated and tested numerically in Refs.~\cite{Jadach:1995hy,Ward:1998ht}.  
The next large contribution to the theoretical error of $\sigma_{\rm ref}$ comes from the vacuum polarisation, more precisely 
from uncertainty of its hadronic part \cite{Jadach:2018jjo,Jadach:2021ayv,Skrzypek:2024gku,Ward:2024frh}.
This, however, should be reduced considerably in the near future due to lattice QCD computations as well as new low-energy
$e^+e^-$ and $e\mu$ experiments, see e.g.\ Ref.~\cite{Aliberti:2025beg,FJ:MTTD2025}.
The other significant uncertainty of the {\sf BHLUMI} predictions comes from the light-fermion pair corrections. 
They, however, can be computed with our Monte Carlo program {\sf KoralW} for all $e^+e^-\rightarrow 4f$ processes 
\cite{Jadach:2001mp}. Although for SABS the $t$-channel $\gamma$-exchange contribution dominates, for the
precision goals of the future $e^+e^-$ colliders, the $s$-channel $\gamma$-exchange as well as $s$ and $t$-channel 
$Z$-exchange contributions (with all their interferences) need to be included. 
All these contributions are implemented in the {\sf BHWIDE} MC generator,
including ${\cal O}(\alpha)$ electroweak corrections in the YFS EEX scheme. Therefore, at this stage our theoretical 
predictions for SABS will be provided with the use of three MC generators 
{\sf BHLUMI} $\oplus$ {\sf BHWIDE}  $\oplus$ {\sf KoralW}.
\vspace{-3mm}
\item{\bf Stage~2:} 
Then, the matrix elements from {\sf BHWIDE} for all $\gamma$ and $Z$-exchange contributions can be implemented
in {\sf BHLUMI}, and similarly for the light-fermion pairs, as described in Refs.~\cite{Jadach:1993wk,Jadach:1996ca}. 
At this stage all the necessary effects will be included in one MCEG -- the ${\cal O}(\alpha^3)_{\rm prag}$ YFS EEX 
version of {\sf BHLUMI}. 
\vspace{-3mm}
\item{\bf Stage~3:} 
Our ultimate goal is to apply the coherent exclusive exponentiation (CEEX)  formalism \cite{Jadach:1998jb}
to SABS, in a similar way as in the MCEG for $e^+\e^-\rightarrow 2f, f\neq e,$ {\sf KKMC} \cite{Jadach:1999vf,Jadach:2022mbe}.
This will result in the ${\cal O}(\alpha^3)_{\rm prag}$ YFS CEEX {\sf BHLUMI} MCEG 
that should satisfy the theoretical precision needs of the luminosity
measurements in the future $e^+e^-$ colliders.
\vspace{-1mm}
\end{description}
In addition to the above main MCEG for SABS, testing tools of {\sf BHWIDE~4.04} will need to be upgraded.
They are necessary for assessing both theoretical and physical precision of the main MCEG. 
They include two MC generators:
the pure leading-log ${\cal O}(\alpha^3L^3)$ {\sf LUMLOG} and the fixed-order ${\cal O}(\alpha)$ {OLDBIS},
as well as the semi-analytical calculations of Ref.~\cite{Jadach:1996bx}. 
For the ${\cal O}(\alpha^3)_{\rm prag}$ precision of {\sf BHLUMI}, the terms $\sim \alpha^4L^4$
would need be included in {\sf LUMLOG}, while {\sf OLDBIS} would need to be upgraded to the fixed-order ${\cal O}(\alpha^2)$ precision,
which should be possible due to the recent progress of two-loop QED calculations with massive leptons \cite{Delto:2023kqv}.
 In the above scheme, the MCEG developed at the earlier stage will also play a role of a testing tool for the MCEG of the next stage.
 
 At LEP, important for establishing the final theoretical precision of the luminosity measurements were cross-checks of {\sf BHLUMI}
 with external MCEGs of a similar physical precision, mainly {\sf SABSPV} \cite{Cacciari:1995fq}. 
 Something similar will be needed for the future $e^+e^-$ colliders, with {\sf BabaYaga} \cite{Balossini:2006wc} 
 as a possible candidate.
 
%================  
\section{LABS with {\sf BHWIDE}}
\label{sec:Bhwide}
%================

Large (or wide) angle Bhabha scattering (LABS), with $\theta_e \gtrsim 100\,$mrad,  is used at the $Z$-peak 
for a direct measurement of the electron partial decay width $\Gamma_e$ of the $Z$ boson \cite{Jadach:1996gu}.
Since the total cross section for a process fermion-pair production the $Z$-peak is
\begin{equation}
\sigma_{e^+e^-\rightarrow 2f}(s = M_Z^2)\:\, \propto\:\, \Gamma_f\Gamma_e\,,
\label{eq:sigma2f}
\end{equation}
LABS is also indirectly used for determination of other fermions partial widths $\Gamma_f$ in a model-independent way.
Measurements of the $Z$-boson partial widths  (or equivalently its decay branching ratios)
are important for precision tests of SM as well as for BSM searches.

At higher electron--positron collision energies, $\sqrt{s} > M_Z$, LABS plays mainly a role of a significant background 
for other processes, in particular di-photon production ($e^+e^-\rightarrow \gamma\gamma$) 
which is also considered as a good candidate for the reference process of 
the luminometry at the future $e^+e^-$ colliders \cite{CarloniCalame:2019dom,deBlas:2024bmz}.   

At lower energies, $\sqrt{s} \lesssim 10\,$GeV, in the so-called flavour factories, LABS is used mainly for
the luminosity measurements \cite{WorkingGrouponRadiativeCorrections:2010bjp}, with the precision at the
 $\sim 0.1\%$ level. 

All the above requires precision theoretical predictions for LABS in the form of a MCEG. 
Most experiments at LEP used {\sf BHWIDE} \cite{Jadach:1995nk,Bhwide:code} as the main MCEG for LABS.
Its precision was estimated at $0.3\%$ near the $Z$-peak (LEP1) and at $1.5\%$ for higher energies (LEP2)
\cite{Jadach:1996gu}, which was sufficient for these experiments. 
For flavour factories (BaBar, Belle, VEPP, BES, KLOE, etc.)
its precision was assessed at ${\cal O}(0.1\%)$ \cite{WorkingGrouponRadiativeCorrections:2010bjp}.
Precision requirements of theoretical predictions for LABS at the future $e^+e^-$ colliders will be much higher than at LEP,
particularly at FCC-ee where a relative experimental precision can reach ${\cal O}(10^{-4})$ at the $Z$ peak and 
${\cal O}(10^{-3})$ at higher energies \cite{deBlas:2024bmz}.

LABS is much more complicated than SABS or other fermion-pair production processes because it involves
both the $\gamma$ and $Z$ exchanges in both the $s$ and $t$ channels, with complex interference patterns
 \cite{Jadach:1996gu}. Therefore, calculations of radiative corrections are more involved for LABS than 
 for the other processes.

The current version of {\sf BHWIDE} \cite{Bhwide:code} features YFS EEX with ${\cal O}(\alpha)$ electroweak (EW) radiative corrections for LABS, 
which means that in addition to the resummation of the QED IR singularities to the infinite order, the non-IR residuals are
calculated up to ${\cal O}(\alpha)$ within SM EW theory. For one-loop virtual corrections, {\sf BHWIDE} is interfaced
with two EW software libraries: {\sf BABAMC} \cite{Bohm:1986fg} and {\sf ALIBABA} \cite{Beenakker:1990mb}. 
The latter goes beyond the strict ${\cal O}(\alpha)$ calculations by using dressed $\gamma$ and $Z$ propagators
through the Dyson resummation of light-fermion contributions to intermediate bosons self-energy corrections. 
This is important because at high energies, these contributions include big logs $\sim\ln(q^2/m_f^2)$,
where $q^2 = s, |t|$ and $m_f$ is the light-fermion mass.  Thus, for the precision $\lesssim 1\%$ at $\sqrt{s} \gtrsim M_Z$,
such a resummation is necessary. For LABS, these self-energy corrections cannot be factorised in terms
of the overall running QED coupling $\alpha(q^2)$, like in SABS, but need to be resummed at the propagator-level
because of the $\gamma$ and $Z$ exchange contributions in both the $s$ and $t$ channels, with all their interferences. 

To satisfy the precision needs of the future high-energy $e^+e^-$ colliders, 
the ${\cal O}(\alpha^2)$ (NNLO) QED corrections will be necessary in {\sf BHWIDE},
and at the $Z$ peak also the NNLO EW corrections (and perhaps ${\cal O}(\alpha^3L^3)$ QED corrections). 
The NNLO QED corrections consist of: (1) the two-loop corrections with massive leptons,  
which have recently been calculated \cite{Delto:2023kqv}, (2) the one-loop corrections to single bremsstrahlung 
with massive leptons, which can be computed with the use of automated software packages, 
such as {\sf OpenLoops} \cite{Buccioni:2019sur} or {\sf Recola} \cite{Actis:2016mpe,Denner:2017wsf},
and (3) the tree-level double-bremsstrahlung matrix element, which can be calculated `by hand' using spin amplitudes
(e.g.\ like for single-bremsstrahlung in the current version of {\sf BHWIDE} \cite{Jadach:1995nk}) 
and/or with the help of some automated software packages.
The NNLO EW corrections at the $Z$ peak can be computed with the {\sf GRIFFIN} software library \cite{Chen:2022dow}. 
Therefore, reaching the precision levels of the future $e^+e^-$ colliders  for LABS by {\sf BHWIDE} looks feasible,
although it will require considerable effort.

%==========================================
\section{Conclusion}
\label{sec:Conc}
%==========================================

In this paper, we have discussed the role of the small and large angle Bhabha scattering processes
and the question of precision needs for their theoretical predictions at the past, current and future 
electron--positron colliders.
We have briefly presented the Monte Carlo event generators {\sf BHLUMI} and {\sf BHWIDE} for these processes,
discussed their current theoretical precision and proposed strategies of their improvements in order to
reach the precision goals of the future $e^+e^-$ colliders.
The theoretical precision of both these generators can be further improved 
by using the recently developed collinearly-enhanced YFS resummation \cite{Jadach:2023pka}.

\section*{Acknowledgments}
We are indebted to late Staszek Jadach -- our mentor, colleague and friend --
for his invaluable contribution to the research described in this paper and for everything we have learnt from him.
This work has been supported by the Polish National Science Centre (NCN) under the grant no.\ 2023/50/A/ST2/00224.

%uncomment the following lines to place a figure
%\begin{figure}[htb]
%\centerline{%
%\includegraphics[width=12.5cm]{Fig1}}
%\caption{Plot of ...}
%\label{Fig:F2H}
%\end{figure}

%\bibliographystyle{IEEEtran}
%\bibliographystyle{unsrt}
%\bibliography{Bhabha}% Produces the bibliography via BibTeX.


\end{document}